\documentclass[12pt,a4paper]{article}
\usepackage{amsmath}
\usepackage{amssymb}

\begin{document}
\bibliographystyle{plain}

\begin{titlepage}

\hfill hep-th/0203184 

\hfill IPPP/02/20

\hfill DCPT/02/40

\vspace{1.5in}

\begin{center}
{\Large{The U(N) ADHM Two-Instanton}} \\
\bigskip
\bigskip
\bigskip
{\large{N.\ B.\ Pomeroy}} \\
\bigskip
{\emph{Department of Physics and IPPP, University of Durham, Durham, DH1 3LE, U.K.}} \\
\bigskip
\texttt{n.b.pomeroy@durham.ac.uk}
\end{center}

\vspace{0.75in}

\begin{center}
{\large{Abstract}} 
\end{center}

The ADHM constraints which implicitly specify instanton gauge field 
configurations are solved for the explicit general form of 
instantons with topological charge two and gauge group $U(N)$. 

\end{titlepage}

Instantons have proved to be an important phenomenon in Yang-Mills field theory ever since their discovery \cite{Belavin:1975fg}. In recent years their use in supersymmetric gauge theories has been considerable. The method for constructing instantons of any given topological charge and for any gauge symmetry was established by Atiyah, Drinfeld, Hitchin and Manin (ADHM) \cite{Atiyah:1978ri}. The ADHM construction of instantons uses techniques from algebraic geometry and the observation by Ward \cite{Ward:1977ta} that gauge fields can be considered as bundles (via a twistor space construction; see also \cite{Atiyah:1977pw,Hartshorne:1978vv}). The ADHM method provides the most general instanton field configurations with the required topological charge. Instantons of higher topological charge represent gauge fields with finite Yang-Mills action which possess increasingly non-trivial topology. However, the equations defining such a configuration contain an amount of redundancy which it is necessary to remove if the instanton configuration so constructed is to be used as a classical solution around which quantum fluctuations occur in the semi-classical approximation method for quantum field theories. The equations resulting from the ADHM construction only implicitly define the instanton configuration, and are referred to as the ADHM constraints. 
In this Letter we solve the ADHM constraints for topological charge of two units and gauge group $U(N)$, for any value of $N$ greater than or equal to two. This constitutes the first exact and explicit general multi-instanton configuration found for the gauge group $U(N)$. 

We begin with a brief exposition of the ADHM construction for the gauge group $U(N)$, closely following the  formalism developed in \cite{Khoze:1998gy}. Unlike the formalism of \cite{Khoze:1998gy}, we shall work in Euclidean space as is conventional for instantons. For the explicit parameterisation of the ADHM matrix $a$, which when in canonical form contains the `ADHM data' from which the instanton gauge field can be constructed, we follow the formalism of \cite{Corrigan:1978ce}. The gauge field $A_{m}(x)$ is an $N \times N$ anti-Hermitian matrix of complex elements, and the gauge field strength is given by $F_{mn} = \partial_{m}A_{n}-\partial_{n}A_{m}+[A_{m},A_{n}]$, in which (as elsewhere) we set the coupling to be $g=1$. (Anti-Hermicity of $A_{m}$ is implemented via $A_{m} \rightarrow iA_{m}$.) The ADHM construction gives the general solution to the self-dual (anti-self-dual) Yang-Mills field equations:
\begin{equation}
F_{mn} = \pm \tfrac{1}{2} \epsilon_{mnkl} F^{kl} = \pm \tilde{F}_{mn}, \label{eq:sd}
\end{equation}
such that the Yang-Mills action assumes the form 
\begin{equation}
k = -\frac{1}{16\pi^{2}}\int \! d^{4}x \: \textup{tr}_{N}(F_{mn} \tilde{F}_{mn}), \label{eq:winnum}
\end{equation}
where $k \in \mathbb{Z}$ is the topological charge, variously known also as the winding number, Pontryagin index, second Chern class, and $\textup{tr}_{N}$ denotes the trace over the gauge group indices. The topological invariant $k$ measures the non-triviality of the topology associated with the mapping which the gauge field $A_{m}$ represents from the field space to the real numbers. As in \cite{Khoze:1998gy}, in our notation an over-bar indicates Hermitian conjugation for matrix quantities, and complex conjugation for scalars quantities.
In the ADHM construction for $U(N)$, one begins with an $(N+2k)\times2k$ complex matrix $\Delta_{[N+2k]\times[2k]}$ (where the subscript notation, introduced in \cite{Khoze:1998gy} indicates the size of the rows and columns of the matrix), which is defined to be linear in the space-time co-ordinate $x$:
\begin{equation}
\Delta(x) \equiv \Delta_{[N+2k]\times[2k]}(x) = a_{[N+2k]\times[k]\times[2]} +b_{[N+2k]\times[k]\times \underline{[2]}}x_{\underline{[2]}\times[2]}. \label{eq:delta1}
\end{equation}
The index $[2k]$ has been decomposed as the direct product of indices $[k]\times[2]$ in order to exhibit the contraction of indices in matrix multiplication: $(AB)_{[a]\times[c]} = A_{[a]\times \underline{[b]}}B_{\underline{[b]}\times[c]}$. The matrices $a$ and $b$ are complex-valued constant matrices which contain the ADHM data describing the instanton, and comprise an overcomplete set of $k$-instanton collective co-ordinates. Here $x$ is represented as a quaternion:
\begin{equation}
x_{[2]\times[2]} = x_{\alpha \dot{\alpha}} = x_{m}\sigma^{m}_{\alpha \dot{\alpha}},
\end{equation}
where $\sigma^{m}_{\alpha \dot{\alpha}}$ are the Euclidean sigma matrices: $\sigma^{m} = (i\tau^{a},1_{[2]\times[2]})_{\alpha \dot{\alpha}}$, with Hermitian conjugate $\bar{\sigma}^{m} = (-i\tau^{a},1_{[2]\times[2]})^{\alpha \dot{\alpha}}$, and $\tau^{a}$ are the standard Pauli matrices ($a=1,2,3$). We can write the quaternionic space-time co-ordinate more explicitly as:
\begin{equation}
x_{\alpha \dot{\alpha}} = x_{m} \sigma^{m}_{\alpha \dot{\alpha}} 
                         =  \left( \! \begin{array}{cc}
                                   x_{4}+ix_{3} & x_{2}+ix_{1} \\
				   -x_{2}+ix_{1} & x_{4}-ix_{3}
				   \end{array} \! \right) 
			= \left( \! \begin{array}{cc}	   
				   z_{1} & z_{2} \\
				   -\bar{z}_{2} & \bar{z}_{1} \end{array} \! \right),
				   \label{eq:xquat}
\end{equation}				   	    
where we have introduced the complex valued co-ordinates $z_{1}$ and $z_{2}$ for later convenience.
The nullspace of the Hermitian conjugate matrix $\bar{\Delta}(x)$, which is $N$-dimensional, has basis vectors which can be contained in an $(N+2k)\times N$-dimensional complex matrix $U(x)$, where:
\begin{equation}
\bar{\Delta}_{[2k]\times\underline{[N+2k]}}U_{\underline{[N+2k]}\times[N]} = \bar{U}_{[N]\times\underline{[N+2k]}}\Delta_{\underline{[N+2k]}\times[2k]} = 0. 
\label{eq:null}
\end{equation}  
The matrix $U(x)$ is orthonormalized to the $N \times N$ unit matrix:
\begin{equation}
\bar{U}_{[N]\times\underline{[N+2k]}}U_{\underline{[N+2k]}\times[N]} = 1_{[N]\times[N]}. 
\end{equation}
The classical (instanton) gauge field $A_m(x)$ can be constructed from the matrix $U(x)$. When the topological charge is zero ($k=0$), the gauge field is given by a gauge transformation of the vacuum (``pure gauge''):
\begin{equation}
A_{m[N]\times[N]} = \bar{U}_{[N]\times\underline{[N+2k]}}\partial_m U_{\underline{[N+2k]}\times[N]}, 
\label{eq:pureg}
\end{equation} 
which automatically satisfies the self-duality equation (Eq.~(\ref{eq:sd})). In the ADHM construction, the condition Eq.~(\ref{eq:pureg}) is taken to give a solution to the self-duality equation for all non-zero values of $k$. This ansatz implies the following factorization condition: 
\begin{equation}
\bar{\Delta}_{[2]\times[k] \times \underline{[N+2k]}} \Delta_{\underline{
[N+2k]} \times[k]\times[2]} = 1_{[2]\times[2]} f^{-1}_{[k]\times[k]}, 
\label{eq:faccon}
\end{equation}
where $f(x)$ is an arbitrary $x$-dependent $k \times k$-dimensional Hermitian matrix. 
When combined with the nullspace condition in Eq.~(\ref{eq:null}), Eq.~(\ref{eq:faccon}) then implies the completeness 
relation:
\begin{equation}
\Delta_{[N+2k]\times\underline{[k]}\times\underline{[2]}} f_{\underline{[k]}\times\underline{[k]}}\bar{\Delta}_{\underline{[2]} \times\underline{[k]} \times[N+2k]} = 1_{[N+2k]\times[N+2k]} - U_{[N+2k]\times\underline{[N]}}\bar{U}_{\underline{[N]}\times[N+2k]}.
\label{eq:comrel}
\end{equation}
Using Eqs.~(\ref{eq:pureg},\ref{eq:faccon},\ref{eq:comrel}) with integration by parts, and using the short-hand notation $X_{[m} Y_{n]} = X_{m} Y_{n} - X_{n} Y_{m}$, the field strength can then be expressed as:
\begin{eqnarray}
F_{mn} & \equiv & \partial_{[m}A_{n]}+A_{[m}A_{n]} = \partial_{[m}(\bar{U}\partial_{n]}U)+(\bar{U}\partial_{[m}U)(\bar{U}\partial_{n]}U) = \partial_{[m}\bar{U}(1-U \bar{U})\partial_{n]}U \nonumber \\ 
& = & \partial_{[m}\bar{U}\Delta f\bar{\Delta}\partial_{n]}U = \bar{U}\partial_{[m}\Delta f \partial_{n]}\bar{\Delta}U = \bar{U} b \sigma_{[m}\bar{\sigma}_{n]}f \bar{b}U = 4\bar{U}b\sigma_{mn}f\bar{b}U,
\end{eqnarray}
where $\sigma_{mn}$ is the numerical tensor defined by $\sigma_{mn} = \frac{1}{4} \sigma_{[m}\bar{\sigma}_{n]}$, which is self-dual, and $\bar{\sigma}_{mn} = \frac{1}{4} \bar{\sigma}_{[m}\sigma_{n]}$, which is anti-self-dual. Due to these properties of $\sigma_{mn}$, the field strength $F_{mn}$ is self-dual or anti-self-dual. We will focus on self-dual instantons. 

The classical (instanton) gauge field so constructed has gauge group $U(N)$. To specify the classical $SU(N)$ instanton gauge field, one can perform a gauge transformation $U \rightarrow Ug_{1}$, where $g_{1} \in U(1)$. 
Continuing with the formalism of \cite{Khoze:1998gy}, we assign the following indices to the objects constituting the ADHM data (the matrices $U$, $\Delta$, $a$, $b$ and $f$, which involve the matrices $\sigma$ and $x$):
\begin{eqnarray*}
\hbox{Instanton number indices\ }[k] & : & 1\le i,j,l \cdots \le k \\
\hbox{Gauge group indices\ }[N] & : & 1 \le u,v \cdots \le N \\
\hbox{ADHM indices\ }[N+2k] & : & 1 \le \lambda,\mu \cdots \le N+2k \\
\hbox{Quaternionic (Weyl) indices\ }[2] & : & \alpha,\beta,\dot{\alpha}, 
\dot{\beta} \cdots=1,2 \\
\hbox{Lorentz indices\ }[4] & : & m,n \cdots=1,2,3,4.
\end{eqnarray*}
No extra notation is required for the $2k$-dimensional column index attached
to $\Delta,$ $a$ and $b$, since it can be factored as $[2k]=[k]\times[2]
=j \dot{\beta},$ etc. In these conventions Eq.~(\ref{eq:delta1}) defining $\Delta(x)$ becomes:
\begin{equation}
\Delta_{\lambda \, i \dot{\alpha}}(x)
\ = \ a_{\lambda \, i \dot{\alpha}}\ +\
b_{\lambda \, i}^{\beta}\,x_{\beta\dot{\alpha}}\ , \qquad
\bar{\Delta}^{\dot{\alpha}\lambda}_{ i  }(x)
\ = \ \bar{a}^{\dot{\alpha}\lambda}_{ i }\ +\
\bar{x}^{\dot{\alpha} \alpha} \, \bar{b}^ \lambda_{\alpha i}, 
\end{equation}
and the factorization condition Eq.~(\ref{eq:faccon}) is
\begin{equation}
\bar{\Delta}^{\dot{\beta}\lambda}_{ j }\,
\Delta_{\lambda \, i \dot{\alpha}} \ = \  \delta^{\dot{\beta}}_{\ \dot{\alpha}} \, (f^{-1})_{ij}. 
\end{equation}
The definition of $\Delta(x)$ and the factorization condition imply the following conditions on $a$ and $b$:
\begin{eqnarray}
\bar{a}^{\dot{\alpha}\lambda}_{ i }\, a_{\lambda  j \dot{\beta}} \
& = & \ (\bar{a} a)_{ij} \delta^{\dot{\alpha}}_{\dot{\beta}} 
\ \propto \ \delta^{\dot{\alpha}}_{\dot{\beta}} \label{eq:adhm1} \\
\bar{a}^{\dot{\alpha}\lambda}_{i}\, b_{\lambda  j}^{\beta} \
& = & \
\bar{b}^{\beta\lambda}_{ i} \, a_{\lambda j}^{\dot{\alpha}} \label{eq:adhm2} \\
\bar{b}^\lambda_{\alpha i} \, b_{\lambda j}^{\beta} \
& = & \ (\bar{b} b)_{ij} \delta_{\alpha}^{\beta} 
\ \propto \ \delta_{\alpha}^{\beta}. \label{eq:adhm3}
\end{eqnarray}
Together, Eqs.~(\ref{eq:adhm1},\ref{eq:adhm2},\ref{eq:adhm3}) constitute the ADHM constraints in their original form. 
The matrices $a$ and $b$ contain the collective co-ordinates of the $k$-instanton gauge field configuration, and the number of these collective co-ordinates increases as $k^{2}$. However, the number of physical collective co-ordinates required to describe the $U(N)$ $k$-instanton is $4Nk$, including global gauge rotations of the gauge field \cite{Bernard:1977nr}. Hence the matrices $a$ and $b$ together form an overcomplete set of collective co-ordinates. Some of the redundancy contained in $a$ and $b$ can be removed via the following $x$-independent transformations under which the ADHM construction is invariant:
\begin{eqnarray}
\Delta_{[N+2k]\times[k]\times[2]} & \rightarrow & \Lambda_{[N+2k]\times \underline{[N+2k]}}\Delta_{\underline{[N+2k]}\times\underline{[k]}\times[2]}B^{-1}_{\underline{[k]}\times[k]}, \nonumber \\
U_{[N+2k]\times[N]} & \rightarrow & \Lambda_{[N+2k]\times\underline{[N+2k]}}U_{\underline{[N+2k]}\times[N]}, \label{eq:adhminv1} \\
f_{[k]\times[k]} & \rightarrow & B_{[k]\times\underline{[k]}}f_{\underline{[k]}\times\underline{[k]}}B^{\dag}_{\underline{[k]}\times[k]}, \nonumber
\end{eqnarray}
where $\Lambda \in U(N+2k)$ and $B \in GL(k, \mathbb{C})$. There are also the usual space-time symmetries associated with the ADHM construction. 
Furthermore, the columns of the matrix $\Delta(x)$ must remain linearly independent in order to avoid singularities in the integrand in Eq.~(\ref{eq:winnum}) \cite{Corrigan:1978ce}.
One can use the symmetries in Eq.~(\ref{eq:adhminv1}) to bring the representation of $a$ and $b$ to the canonical form \cite{Corrigan:1978ce} by removing the degrees of freedom from the matrix $b$:
\begin{equation}
a_{[N+2k]\times[2k]} = \left( \! \begin{array}{c}
                              u_{[N] \times[2k]} \\ a^{\prime}_{[2k]\times[2k]} \end{array} \! \right), 
			      \qquad 
b_{[N+2k]\times[2k]} = \left( \! \begin{array}{c} 
 			       0_{[N] \times[2k]} \\ 1_{[2k]\times[2k]} \end{array} \! \right).
\label{eq:canon}
\end{equation}			       
The sub-matrix elements $a^{\prime} \equiv (a^{\prime}_{\alpha \dot{\alpha}})_{ij}$ are also represented using a quaternionic basis: 
\begin{equation}
(a^{\prime}_{\alpha \dot{\alpha}})_{ij} = (a^{\prime}_{n})_{ij} \sigma_{n \alpha \dot{\alpha}}, \qquad 
(\bar{a}^{\prime \alpha \dot{\alpha}})_{ij} = (a^{\prime}_{n})_{ij}\bar{\sigma}^{\alpha \dot{\alpha}}_{n},
\end{equation}
whilst the sub-matrix $u$ is a complex-valued matrix (which can also be represented as a quaternion).
In addition to invariance under the transformations in Eq.~(\ref{eq:adhminv1}) there exists an auxiliary, or residual, symmetry arising from the symmetry of the ADHM construction in Eq.~(\ref{eq:adhminv1}): the canonical form of $b$ given in Eq.~(\ref{eq:canon}) is invariant under global $U(k) \in U(N+2k) \times GL(k, \mathbb{C})$ rotations, which act upon $\Delta_{[N+2k]\times[2k]}$ as  
\begin{equation}
\Delta_{[N+2k]\times[2k]} \rightarrow \left( \! \begin{array}{cc}
1_{[N]\times[N]} & 0_{[2k]\times[N]} \\
0_{[N]\times[2k]} & \bar{\Lambda}_{[2k]\times[2k]} \end{array} \! \right) \Delta_{[N+2k]\times[2k]}\Lambda_{[2k]\times[2k]},
\label{eq:adhminv2}
\end{equation}
where $\Lambda_{[2k]\times[2k]} = \Omega_{[k]\times[k]}1_{[2]\times[2]}$ and $\Omega_{[k]\times[k]} \in U(k)$. This auxiliary $U(k)$ symmetry can be employed to simplify the final form of solutions of the ADHM constraints.
With $a$ and $b$ in the canonical form, the ADHM constraint Eq.~(\ref{eq:adhm3}) is automatically satisfied, and the other two constraints give:
\begin{eqnarray}
\textup{tr}_{2}(\tau^{a \dot{\alpha}}_{\dot{\beta}} \bar{a}^{\dot{\beta}} a_{\dot{\alpha}}) & = & 0, \label{eq:adhmv1} \\
(a^{\prime}_{n})^{\dag}_{ij} & = & (a^{\prime}_{n})_{ij}.
\label{eq:adhmv2}
\end{eqnarray}
The three Pauli matrices $\tau^{a}$ have been used to contract the product $(\bar{a}a)$, ($\textup{tr}_{2}$ denotes the trace over quaternionic (Weyl) indices), so that the above constraint Eq.~(\ref{eq:adhmv1}) gives three distinct equations. It is at this point that we now transfer to the formalism used by Corrigan et al.\ in \cite{Corrigan:1978ce} for a simple parameterisation of the ADHM matrix $a$. Since we are dealing with instantons in $U(N)$, we point out that it is not necessary to use quaternions to derive the ADHM constraints in this case; complex-valued matrices can be used instead. For our purposes, we now revert to using complex-valued matrices. In doing this, we have the following `dictionary' for translating the above formalism of \cite{Khoze:1998gy} to that of \cite{Corrigan:1978ce}:
\begin{eqnarray}
a_{[N+2k]\times[2k]} & = & \left( \! \begin{array}{c}
                               u_{[N]\times[2k]} \\ a^{\prime}_{[2k]\times[2k]} \end{array} \! \right)
		     =	\left( \! \begin{array}{cc}
		               u_{1[N]\times[k]} &  u_{2[N]\times[k]}  \\  
			       (a^{\prime}_{4}+ia^{\prime}_{3})_{[k]\times[k]} & (a^{\prime}_{2}+ia^{\prime}_{1})_{[k]\times[k]} \\
			       (-a^{\prime}_{2}+ia^{\prime}_{1})_{[k]\times[k]} & (a^{\prime}_{4}-ia^{\prime}_{3})_{[k]\times[k]} \end{array} \! \right) \nonumber \\
	             & \equiv & \left( \! \begin{array}{cc}
		               u_{1[N]\times[k]} &  u_{2[N]\times[k]}  \\
			       r_{11[k]\times[k]} & r_{12[k]\times[k]} \\
			       -\bar{r}_{12[k]\times[k]} & \bar{r}_{11[k]\times[k]} \end{array} \! \right), 
\label{eq:cftg}
\end{eqnarray}			      
where we have introduced the complex-valued matrices $u_{1}$, $u_{2}$, $r_{11}$ and $r_{12}$. The canonical form of $b$ in Eq.~(\ref{eq:canon}) remains unchanged. We label the elements of $u_{\alpha}$ as $u_{\alpha,uv}$, using the same index assignments given above and use the same index placement for the elements $u_{\alpha,uv}$ upon complex conjugation (e.g.\ for $u_{1,11}$, we have $u^{\ast}_{1,11} \equiv \bar{u}_{11,1}$) as in \cite{Khoze:1998gy}. After specialising to $k=2$, we will deal with the elements of the sub-matrices $r_{11}$ and $r_{12}$ explicitly. The formalism \cite{Corrigan:1978ce} now employed simplifies the form of the sub-matrix $a^{\prime}$, and consequently the form of the ADHM constraints, as now $a^{\prime}$ involves only two complex-valued matrices ($r_{11}$ and $r_{12}$), instead of four Hermitian matrices (that is, the matrices $(a^{\prime}_{n})_{ij}$) in the previous formulation \cite{Khoze:1998gy}.  

The $U(N)$ $k$-instanton ADHM constraints can be expressed in terms of the complex-valued matrices $r_{11}$, $r_{12}$, $u_{1}$ and $u_{2}$ as:
\begin{eqnarray}
\bar{u}_{1}u_{2} + \bar{r}_{11}r_{12} - r_{12}\bar{r}_{11} & = & 0, \label{eq:cftg1} \\
\bar{u}_{1}u_{1} - \bar{u}_{2}u_{2} + \bar{r}_{11}r_{11} - r_{11}\bar{r}_{11} + r_{12}\bar{r}_{12} - \bar{r}_{12}r_{12} & = & 0.
\label{eq:cftg2}
\end{eqnarray}
The first constraint, Eq.~(\ref{eq:cftg1}), and second constraint, Eq.~(\ref{eq:cftg2}), respectively, are often referred to as the `complex ADHM constraint,' and the `real ADHM constraint.' For $k \ge 2$, however, both of these matrix equations will contain real and complex elements. Note that there is no analogue of the ADHM constraint Eq.~(\ref{eq:adhmv2}) in this formulation; this is because both the real and the complex parts of the sub-matrix $a^{\prime}$ are contained in $r_{11}$ and $r_{12}$. Consequently, the only remaining constraints, Eq.~(\ref{eq:adhmv1}), give the fundamental $U(N)$ $k$-instanton ADHM constraints Eqs.~(\ref{eq:cftg1},\ref{eq:cftg2}), and Eq.~(\ref{eq:adhmv2}) is a remnant of formalism from \cite{Khoze:1998gy}. 

We can now count the number of real independent parameters which solutions of the ADHM constraints shall possess. The ADHM matrix $a$, in the form given in Eq.~(\ref{eq:cftg}) contains $4Nk+4k^{2}$ real parameters. The ADHM constraints in the formalism of \cite{Corrigan:1978ce}, Eqs.~(\ref{eq:cftg2}), then place $3k^{2}$ real conditions on the elements of $a$. The auxiliary $U(k)$ symmetry removes a further $k^{2}$ real parameters, giving a total of:
\begin{equation}
4Nk+4k^{2}-3k^{2}-k^{2} = 4Nk,
\label{eq:fornk}
\end{equation}
which includes the global gauge rotations implemented by the auxiliary $U(k)$ symmetry. To obtain a solution with purely physical degrees of freedom, that is, only the true collective co-ordinates required to specify the position, size and iso-orientation (in group space) of the $k$-instanton, the global gauge rotations are removed, leaving the number of independent physical parameters as $4Nk-N^{2}+1$ for $ k \ge \frac{1}{2}N$, and $4k^{2}+1$ for $k \le \frac{1}{2}N$ \cite{Corrigan:1978ce,Bernard:1977nr}. For the purposes of using ADHM instanton configurations in instanton calculus, however, the global gauge rotations must be included, since they appear in the $k$-instanton measure and are to be integrated over \cite{Osborn:1981yf,Dorey:1996hu}. The ADHM constraints are both bilinear and quadratic in the elements of the sub-matrices $\{ u_{1}, u_{2}, r_{11}, r_{12} \}$, and also couple these elements to each other. As the topological charge $k$ increases, the number of elements in these sub-matrices increases, and the complexity of the constraints increase. Essentially, the ADHM constraints present the problem of constrained, coupled, non-linear simultaneous algebraic equations in complex or quaternionic variables, or more generally, in matrix variables.

The ADHM constraints contain a great deal of complexity, which the appearance of Eqs.~(\ref{eq:cftg1},\ref{eq:cftg2}) does not make immediately apparent. An indication of this complexity is that, since the inception of the ADHM construction \cite{Atiyah:1978ri}, there have only been a small number of exact three-instanton configurations found, \cite{Christ:1978jy,Korepin:1983bn} amongst them, all of which are for the gauge group $Sp(1) \simeq SU(2)$ and have twenty-one parameters. No exact completely general instanton configurations of topological charge equal to or higher than three have been determined, although methods for obtaining $Sp(1) \simeq SU(2)$ four-instanton configurations have been proposed \cite{Korepin:1983bn,Inozemtsev:1982av}. Pioneering $k$-instanton configurations have been found, for example in \cite{Jackiw:1976dw}, but these do not constitute general solutions of the ADHM constraints due to the number of parameters which describe them. An $SU(N)$ $k$-instanton configuration with $N=2k$ has been constructed within the ADHM method \cite{Drinfeld:1978xr}, but again this is a special (though exact) solution of the ADHM constraints. 

With the aim of determining the most general solution of these constraints for $k=2$, which will have $8N$ real independent parameters, we firstly need to find a solution with $8N+4$ real parameters. This then allows one to rotate out the $U(2)$ symmetry, effectively eliminating four real parameters in the $8N+4$-parameter solution. To simplify notation, we define the following quantities in terms of the ADHM data present in the matrix $a$:
\begin{equation}
\bar{u}_{1}u_{2} = \sum^{N}_{n=1} \left( \! \begin{array}{cc}
                                                \bar{u}_{n1,1}u_{2,n1} & \bar{u}_{n1,1}u_{2,n2} \\
						\bar{u}_{n2,1}u_{2,n1} & \bar{u}_{n2,1}u_{2,n2} \end{array} \! \right)
						\equiv \left( \! \begin{array}{cc} 
						                   U_{x} & U_{y} \\
								   U_{z} & U_{t} \end{array} \! \right),
\label{eq:bux}
\end{equation}
\begin{equation}
\bar{u}_{1}u_{1} - \bar{u}_{2}u_{2} = \sum^{N}_{n=1} \left( \! \begin{array}{cc}
                                                                |u_{1,n1}|^{2} - |u_{2,n1}|^{2} & \bar{u}_{n1,1}u_{1,n2} - \bar{u}_{n1,2}u_{2,n2} \\
u_{n1,1}\bar{u}_{1,n2} - u_{n1,2}\bar{u}_{2,n2}	 &  |u_{1,n2}|^{2} - |u_{2,n2}|^{2} \end{array} \! \right) \equiv \left( \! \begin{array}{cc} U_{1} & U_{2} \\
                         \bar{U}_{2} & U_{4} \end{array} \! \right). 
\label{eq:bu1}
\end{equation}
Note that the sums in Eqs.~(\ref{eq:bux},\ref{eq:bu1}) run from $1$ to $N$; although the number $N$ in these sums is related to the rank of the gauge group $U(N)$, the construction of ADHM instanton configurations breaks down for $N=1$, as is to be expected given that (commutative) ADHM instantons are a phenomenon of non-Abelian gauge theories.
We now make a change of variables which affects only the diagonal elements of the matrices $r_{11}$ and $r_{12}$, such that:
\begin{eqnarray}
r_{11} & = & \left( \! \begin{array}{cc} a+\frac{1}{2}x_{0} & b \\ c & a-\frac{1}{2}x_{0} \end{array} \! \right), \\ 
r_{12} & = & \left( \! \begin{array}{cc} \alpha+\frac{1}{2}x_{1} & \beta \\ \gamma & \alpha-\frac{1}{2}x_{1} \end{array} \! \right),
\label{eq:rexp}
\end{eqnarray}
where $\{ x_{0},x_{1},a,b,c,\alpha,\beta,\gamma \} \in \mathbb{C}$. This is to make the physical interpretation of the $U(N)$ $k=2$ ADHM instanton configurations more transparent, and to simplify calculations involving the elements of $r_{11}$ and $r_{12}$.
Our main result is the solution of the $U(N)$ $k=2$ ADHM constraints with $8N+4$ real parameters, which is:
\begin{eqnarray}
r_{11} & = & \left( \! \begin{array}{cc} 
                     a+\frac{1}{2}x_{0} & \frac{\displaystyle{x_{0}[\bar{P}u-P]}}{\displaystyle{|x|^{2}|x_{1}|^{2}(|u|^{2}-1)}}-\frac{\displaystyle{\bar{U}_{z}}}{\displaystyle{\bar{x}_{1}}} 
\label{eq:un2sol1}		     
\\
		     \frac{\displaystyle{\bar{U}_{y}}}{\displaystyle{\bar{x}_{1}}}-\frac{\displaystyle{x_{0}\bar{U}_{x}}}{\displaystyle{\bar{U}_{z}}}-\frac{\displaystyle{x_{0}\bar{u}[\bar{P}u-P]}}{\displaystyle{|x|^{2}|x_{1}|^{2}(|u|^{2}-1)}} & a-\frac{1}{2}x_{0} \end{array} \! \right), \\ \nonumber \\
r_{12} & = & \left( \! \begin{array}{cc} 
                     \alpha+\frac{1}{2}x_{1} & -\frac{\displaystyle{x_{1} \bar{U}_{x}}}{\displaystyle{{U}_{z}}}-\frac{\displaystyle{x_{1}u[P\bar{u}-\bar{P}]}}{\displaystyle{|x|^{2}|x_{1}|^{2}(|u|^{2}-1)}} \\
		     \frac{\displaystyle{x_{1}[P\bar{u}-\bar{P}]}}{\displaystyle{|x|^{2}|x_{1}|^{2}(|u|^{2}-1)}} & \alpha-\frac{1}{2}x_{1} \end{array} \! \right), 
\label{eq:un2sol2}	    
\end{eqnarray}
\\
where 
\begin{eqnarray}
P & \equiv & \bar{x}_{0}x_{1}\bar{U}_{z}+x_{0}\bar{x}_{1}U_{y}-|x_{1}|^{2}U_{2}-|x|^{2}|x_{1}|^{2}\frac{U_{x}}{U_{y}}, \nonumber \\
u & \equiv & \frac{U_{y}}{U_{z}}, \\
|x|^{2} & = & |x_{0}|^{2} - |x_{1}|^{2}, \nonumber
\end{eqnarray}
and includes the following conditions which also arise from the constraints (Eqs.~(\ref{eq:cftg1},\ref{eq:cftg2})), which when taken together with Eqs.~(\ref{eq:un2sol1},\ref{eq:un2sol2}) constitute the general solution of the $U(N)$ $k=2$ ADHM constraints:
\begin{eqnarray}
U_{x} & = & -U_{t}, \label{eq:trace1} \\
U_{1} & = & -U_{4} \ = \ |b|^{2}-|c|^{2}+|\gamma|^{2}-|\beta|^{2}, \label{eq:trace2}
\end{eqnarray}
Equation (\ref{eq:trace1}) and the first equality in Eq.~(\ref{eq:trace2}) can be obtained from taking the trace over the instanton number index of the ADHM constraints (Eqs.~(\ref{eq:cftg1},\ref{eq:cftg2})); the second equality in Eq.~(\ref{eq:trace2}) occurs in the second ADHM constraint, Eq.~(\ref{eq:cftg2}). 
There remains the aforementioned $U(k)$ auxiliary symmetry in the solution in Eqs.~(\ref{eq:un2sol1}--\ref{eq:trace2}), the removal of which will reduce the number of real, independent physical parameters to $8N$. The $U(k)$ symmetry of Eq.~(\ref{eq:adhminv2}) acts as follows on the sub-matrices of $a$:
\begin{eqnarray}
\left( \! \begin{array}{lr} u_{1} & u_{2} \end{array} \! \right) & \rightarrow & \left( \! \begin{array}{lr} u_{1}\Omega & u_{2}\Omega \end{array} \! \right), \\
\left( \! \begin{array}{cc} r_{11} & r_{12} \\ -\bar{r}_{12} & \bar{r}_{11} \end{array} \! \right) & \rightarrow & 
\Omega^{\dag} \left( \! \begin{array}{cc} r_{11} & r_{12} \\ -\bar{r}_{12} & \bar{r}_{11} \end{array} \! \right) \Omega, \label{eq:aux}
\end{eqnarray}
where $\Omega \in U(2)$ for topological charge $k=2$. In the Appendix we provide the details of a particular transformation $\Omega$ which can be used to set $U_{x}=0$ and $u_{1,11}=0$ (indeed any other element of $u_{1}$ or $u_{2}$ could be set to zero). We adopt this usage of the $U(2)$ auxiliary symmetry hereon.  
We note that any other solutions of the ADHM constraints for gauge group $U(N)$ and $k=2$ possessing $8N+4$ real parameters will be equivalent to the above solution in Eqs.~(\ref{eq:un2sol1}--\ref{eq:trace2}) upon acting on it with the auxiliary $U(2)$ symmetry, since the instanton moduli space for $k=2$ is connected \cite{Hartshorne:1978vv}. 

Using the $U(2)$ transformation permits one to construct the $U(N)$ $k=2$ ADHM instanton which has a definite physical interpretation. The form of the $8N$-parameter solution, following on from the $8N+4$-parameter solution Eqs.~(\ref{eq:un2sol1}--\ref{eq:trace2}) is then: 
\begin{eqnarray}
r_{11} & = & \left( \! \begin{array}{cc} 
                     a+\frac{1}{2}x_{0} & \frac{\displaystyle{x_{0}[\bar{R}u-R]}}{\displaystyle{|x|^{2}|x_{1}|^{2}(|u|^{2}-1)}}-\frac{\displaystyle{\bar{U}_{z}}}{\displaystyle{\bar{x}_{1}}} 
\label{eq:un2p1}		     
\\
		     \frac{\displaystyle{\bar{U}_{y}}}{\displaystyle{\bar{x}_{1}}}-\frac{\displaystyle{x_{0}\bar{u}[\bar{R}u-R]}}{\displaystyle{|x|^{2}|x_{1}|^{2}(|u|^{2}-1)}} & a-\frac{1}{2}x_{0} \end{array} \! \right), \\
r_{12} & = & \left( \! \begin{array}{cc} 
                     \alpha+\frac{1}{2}x_{1} & -\frac{\displaystyle{x_{1}u[R\bar{u}-\bar{R}]}}{\displaystyle{|x|^{2}|x_{1}|^{2}(|u|^{2}-1)}} \\
		     \frac{\displaystyle{x_{1}[R\bar{u}-\bar{R}]}}{\displaystyle{|x|^{2}|x_{1}|^{2}(|u|^{2}-1)}} & \alpha-\frac{1}{2}x_{1} \end{array} \! \right),
\label{eq:un2p2}
\end{eqnarray}
where $R$ is the function
\begin{equation}
R \equiv \bar{x}_{0}x_{1}\bar{U}_{z}+x_{0}\bar{x}_{1}U_{y}-|x_{1}|^{2}U_{2}, 
\label{eq:rdef}
\end{equation}
and the conditions Eq.~(\ref{eq:trace1},\ref{eq:trace2}) (with $U_{x}=0$ and $u_{1,11}=0$) complete the specification of the $8N$-parameter $U(N)$ $k=2$ ADHM instanton solution. 
For the case $N=2$, the explicit two-instanton configuration can assume a particularly simple form. Continuing with our choice of $U(2)$ symmetry used as $U_{x}=0$ and $u_{1,11}=0$, for the $U(2)$ two-instanton one has:
\begin{eqnarray}
U_{x} & = & \bar{u}_{1,21}u_{2,21} = 0, \\ \label{eq:u2rels1}
U_{y} & = & \bar{u}_{1,21}u_{2,22}, \\
U_{z} & = & \bar{u}_{1,12}u_{2,11}+\bar{u}_{1,22}u_{2,21}, \\
U_{t} & = & \bar{u}_{1,12}u_{2,12}+\bar{u}_{1,22}u_{2,22}, 
\end{eqnarray}
and if further one takes $u_{1,21}=0$, then $U_{y}=0$, $U_{z}$ and $U_{t}$ remain unmodified, and 
\begin{eqnarray}
U_{2} & = & -(\bar{u}_{2,11}u_{2,12}+\bar{u}_{2,21}u_{2,22}), \\
U_{1} & = & -|u_{2,11}|^{2}-|u_{2,21}|^{2}. \label{eq:u2rels2}
\end{eqnarray}
With these choices, any off-diagonal elements proportional to $U_{y}$ then vanish, and the matrices $r_{11}$ and $r_{12}$ in Eqs.~(\ref{eq:un2p1},\ref{eq:un2p2}) for $N=2$ simplify to:
\begin{eqnarray}
r_{11} & = & \left( \! \begin{array}{cc} a+\frac{1}{2}x_{0} & \frac{\displaystyle{1}}{\displaystyle{|x|^{2}|x_{1}|^{2}}}[x_{1}|x_{0}|^{2}\bar{U}_{z}-x_{0}|x_{1}|^{2}U_{2}] \\ 0 & a-\frac{1}{2}x_{0} \end{array} \! \right), \\ 
r_{12} & = & \left( \! \begin{array}{cc} \alpha+\frac{1}{2}x_{1} & 0 \\ \frac{\displaystyle{1}}{\displaystyle{|x|^{2}|x_{1}|^{2}}}[x_{0}|x_{0}|^{2}U_{z}-x_{1}|x_{1}|^{2}\bar{U}_{2}] & \alpha-\frac{1}{2}x_{1} \end{array} \! \right),
\end{eqnarray}
Using Eqs.~(\ref{eq:u2rels1}--\ref{eq:u2rels2}) and the choice $u_{1,21}=0$, one can choose to eliminate $u_{1,12}$ via the relation $U_{x}=-U_{t}=0$, (Eq.~(\ref{eq:trace1})), making $U_{z}$ proportional to $\bar{u}_{1,22}$. The modulus of $u_{1,22}$ can thus be eliminated via $U_{1}=-U_{4}$, (Eq.~(\ref{eq:trace2})), and the remaining constraint, the second equality in Eq.~(\ref{eq:trace2}), enables one to eliminate the imaginary part of $u_{1,22}$ through a quadratic relation in this quantity. A similar procedure has been performed for $N=3$, in which the constraint Eq.~(\ref{eq:trace2}) becomes more involved, but other choices of elements within $u_{1}$ and $u_{2}$ to eliminate could be chosen to simplify this. The number of independent real parameters remaining in the solution is then sixteen (eight from $\{ u_{2,11}, u_{2,12}, u_{2,21}, u_{2,22} \}$ and eight from $\{ a, \alpha, x_{0},x_{1}\}$), which agrees with the general result of $8N=16$ real parameters from the parameter counting in Eq.~(\ref{eq:fornk}). Hence, upon fixing the auxiliary $U(2)$ symmetry, the above ADHM data for the $U(2)$ two-instanton configuration represents the unique sixteen parameter solution of the ADHM constraints for the gauge group $U(2)$ and topological charge $k=2$. 
We note that physical quantities constructed from the $SU(2)$ two-instanton configuration, which can be obtained from the $U(2)$ two-instanton configuration given above, will be equivalent to those constructed from the $Sp(1)$ two-instanton \cite{Christ:1978jy} due to the isomorphism $SU(2) \simeq Sp(1)$.  

In the case of gauge group $U(N)$, with $N>1$, the following indentification of physical parameters in this solution can be made. The instanton centre of mass co-ordinates (translational co-ordinates) are given by $a$ and $\alpha$, which are proportional to $x_{m}$ and can thus be set to zero. The relative instanton positions are then taken to be $x_{0}$ and $x_{1}$. The scale sizes can be expressed using the definition given for the $U(N)$ $k$-instanton scale sizes in \cite{Dorey:1999pd} as:
\begin{eqnarray}
\rho^{2}_{1} & = & \tfrac{1}{2}U_{4} - \tfrac{1}{2}\sum^{N}_{n=1}|u_{2,n2}|^{2} = \tfrac{1}{2}\sum^{N}_{n=1}(|u_{1,n1}|^{2}-|u_{2,n1}|^{2}), \label{eq:scale1} \\
\rho^{2}_{2} & = & \tfrac{1}{2}U_{1} - \tfrac{1}{2}\sum^{N}_{n=1}|u_{2,n1}|^{2} = 
\tfrac{1}{2}\sum^{N}_{n=1}(|u_{1,n2}|^{2}-|u_{2,n2}|^{2}). \label{eq:scale2}
\end{eqnarray}
The global gauge orientations (which will include iso-orientations for any chosen $N$) are given by the remaining parameters contained within the sub-matrices $u_{1}$ and $u_{2}$, as they serve to rotate the two-instanton solution in the group space of $U(N)$; through these sub-matrices any $U(N)$ two-instanton can be specified, and no embedding is necessary at this stage. 
Thus, for the $U(2)$ solution given above, the relative instanton separations are $\{ x_{0}, x_{1} \}$, the instanton centre of mass positions are $\{ a, \alpha \}$, and the two scale sizes are $\rho_{1}$ and $\rho_{2}$, as defined in Eqs.~(\ref{eq:scale1},\ref{eq:scale2}). The six $U(2)$ iso-orientations are contained in the remaining elements $\{ u_{2,11}, u_{2,12}, u_{2,21}, u_{2,22} \}$ taken together with the conditions which relate them.  

We can now make a count of the parameters appearing in the $U(N)$ two-instanton solution. The instanton translational co-ordinates and relative separations, $\{ a, \alpha, x_{0},x_{1}\}$, give eight real parameters. There are two scale sizes, $\{ \rho_{1}, \rho_{2} \}$, given by Eqs.~(\ref{eq:scale1},\ref{eq:scale2}), which are two real parameters. Also there are $(4N-5)k = 8N-10$ real iso-orientations. Summing these gives $8N-10+8+2 = 8N$ real parameters, as required by the parameter counting in Eq.~(\ref{eq:fornk}).
This solution must also exhibit the correct decomposition into two constituent one-instanton configurations in the dilute instanton gas limit, which is a physically required property. This asymptotic limit can most simply be achieved by taking the relative instanton positions to infinity; that is, the separation of the two coupled one-instantons approximately comprising the two-instanton is taken to be infinite in extent. In this way, the description of the two-instanton is approximated as a non-interacting gas of two one-instantons (`single instantons'). The result is that the matrix $a$ can be explicitly decomposed for $N=2$ as:
\begin{equation}
a_{[6]\times[4]} \rightarrow \left( \! \begin{array}{cccc} 
u_{1,11} & 0 & u_{2,11} & 0 \\
u_{1,21} & 0 & u_{2,21} & 0 \\
a+\frac{1}{2}x_{0} & 0 & \alpha+\frac{1}{2}x_{1} & 0 \\
0 & 0 & 0 & 0 \\
-\bar{\alpha}-\frac{1}{2}\bar{x}_{1} & 0 & \bar{a}+\frac{1}{2}\bar{x}_{0} & 0 \\
0 & 0 & 0 & 0 
\end{array} \! \right) + 
\left( \! \begin{array}{cccc} 
0 & u_{1,12} & 0 & u_{2,12} \\
0 & u_{1,22} & 0 & u_{2,22} \\
0 & 0 & 0 & 0 \\
0 & a-\frac{1}{2}x_{0} & 0 & \alpha-\frac{1}{2}x_{1} \\
0 & 0 & 0 & 0 \\
0 & -\bar{\alpha}+\frac{1}{2}\bar{x}_{1} & 0 & \bar{a}-\frac{1}{2}\bar{x}_{0} 
\end{array} \! \right), \label{eq:limit}
\end{equation}
where we have restored $u_{1,11}$ and $u_{1,21}$ for clarity (other choices of $U(2)$ transformation can be used).
The one-instantons are centered at $(x_{0},x_{1})$ and at $(-x_{0},-x_{1})$, respectively, and have scale sizes $\rho_{1}$ and $\rho_{2}$. This decomposition in the dilute instanton gas limit can be extended to $U(N)$, in which case the sub-matrix $a^{\prime}$ will decompose in the same manner as in Eq.~(\ref{eq:limit}) and the sub-matrices $u_{1}$ and $u_{2}$ will decompose in a similar way. The $U(2)$ two-instanton may perhaps assist in uncovering a `dictionary' relating it to the $Sp(1)$ ADHM formalism \cite{Khoze:1998gy}, thus connecting the collective co-ordinates which describe these instantons. 
 
We now outline the construction of the instanton gauge field $A_{m}$ and indicate how the gauge field for $U(N)$ with $k=2$ can be obtained from our $8N$-parameter solution given by Eqs.~(\ref{eq:trace1},\ref{eq:trace2},\ref{eq:un2p1}--\ref{eq:rdef}). As in the formalism of \cite{Khoze:1998gy}, we adopt the following decomposition for the ADHM objects $\Delta$ and $U$:
\begin{equation}
U_{[N+2k]\times[N]} = \left( \! \begin{array}{cc} V_{[N]\times[N]} \\ U^{\prime}_{[2k]\times[N]} \end{array} \! \right), \ \Delta_{[N+2k]\times[2k]} = \left( \! \begin{array}{cc} u_{[N]\times[2k]} \\ \Delta^{\prime}_{[2k]\times[2k]} \end{array} \! \right).
\label{eq:decomp}
\end{equation}
One can construct $A_{m}$ by determining $U$ in terms of $\Delta$, which contains the resolved instanton configuration specified by $a$ and $b$. Substituting Eq.~(\ref{eq:decomp}) into the completeness relation Eq.~(\ref{eq:comrel}), one obtains 
\begin{equation}
V_{[N]\times\underline{[N]}}\bar{V}_{\underline{[N]}\times[N]} = 1_{[N]\times[N]}-u_{[N]\times\underline{[k]}\times\underline{[2]}}f_{\underline{[k]}\times\underline{[k]}}\bar{u}_{\underline{[2]}\times\underline{[k]}\times[N]}.
\label{eq:vvbar}
\end{equation}
Any matrices $V$ which solve Eq.~(\ref{eq:vvbar}) are related to each other by the gauge transformation $V \rightarrow Vg_{N}$, where $g_{N} \in U(N)$. Selecting a particular $V$ is thus associated with fixing the (local) gauge of the instanton. Following \cite{Khoze:1998gy}, we choose to work in `singular gauge,' in which $V$ is given by one of the matrix square roots of Eq.~(\ref{eq:vvbar}):
\begin{equation}
V = (1-uf\bar{u})^{1/2}.
\label{eq:vroot}
\end{equation}
Again using Eq.~(\ref{eq:comrel}) the matrix $U^{\prime}$ can then be expressed in terms of $V$ as:
\begin{equation}
U^{\prime} = - \Delta^{\prime}f\bar{u}\bar{V}^{-1}.
\label{eq:uprim}
\end{equation}
To initiate this procedure, one first determines the $x$-dependent Hermitian matrix $f_{[k]\times[k]}$. For $k=2$, we use Eqs.~(\ref{eq:delta1},\ref{eq:xquat}) for constructing the matrix $\Delta = a+bx$. Then the factorisation condition Eq.~(\ref{eq:faccon}) in terms of the elements belonging to the matrices $r_{11}$, $r_{12}$ (of Eqs.~(\ref{eq:un2p1},\ref{eq:un2p2})) and $u_{1},u_{2}$ is:
\begin{equation}
\bar{\Delta}\Delta = f^{-1}1_{[2]\times[2]} = \left( \! \begin{array}{cc} f_{1}^{-1} & 0_{[2]\times[2]} \\ 0_{[2]\times[2]} & f_{2}^{-1} \end{array} \! \right),
\label{eq:exfaccon}
\end{equation}
where 
\begin{eqnarray*}
f_{1}^{-1} = \sum^{N}_{n=1} \left( \! \begin{array}{cc} |u_{1,n1}|^{2}+|A_{1}|^{2}+|B_{1}|^{2}+|c|^{2}+|\beta|^{2} & \bar{u}_{n1,1}u_{1,n2}+\bar{A}_{1}b+A_{2}\bar{c}+
B_{1}\bar{\gamma}+\bar{B}_{2}\beta \\
u_{1,n1}\bar{u}_{n2,1}+A_{1}\bar{b}+\bar{A}_{2}c+
\bar{B}_{1}\gamma+B_{2}\bar{\beta} & |u_{1,n2}|^{2}+|A_{2}|^{2}+|B_{2}|^{2}+|b|^{2}+|\gamma|^{2}
\end{array} \! \right), \\
f_{2}^{-1} = \sum^{N}_{n=1} \left( \! \begin{array}{cc} |u_{2,n1}|^{2}+|A_{1}|^{2}+|B_{2}|^{2}+|b|^{2}+|\gamma|^{2} & \bar{u}_{n1,2}u_{2,n2}+A_{1}\bar{c}+\bar{A}_{2}b+
\bar{B}_{1}\beta+B_{2}\bar{\gamma} \\
u_{2,n1}\bar{u}_{n2,2}+\bar{A}_{1}c+A_{2}\bar{b}+
B_{1}\bar{\beta}+\bar{B}_{2}\gamma & |u_{2,n2}|^{2}+|A_{2}|^{2}+|B_{2}|^{2}+|c|^{2}+|\beta|^{2}
\end{array} \! \right),
\end{eqnarray*}
in which we have defined
\begin{equation}
A_{1} \equiv a+\tfrac{1}{2}x_{0}+z_{1}, \hspace{0.08in} A_{2} \equiv a-\tfrac{1}{2}x_{0}+z_{1}, \hspace{0.08in} 
B_{1} \equiv \alpha+\tfrac{1}{2}x_{1}+z_{2},  \hspace{0.08in} B_{2} \equiv \alpha-\tfrac{1}{2}{x}_{1}+z_{2}. 
\label{eq:newvar}
\end{equation} 
We note that the product $f^{-1}1_{[2]\times[2]}$ in Eq.~(\ref{eq:exfaccon}) is a possible source of some ambiguity. The correct form of this product becomes clear when compared with the result of calculating $\bar{\Delta}\Delta$.
One can choose to invert either $f_{1}^{-1}$ or $f_{2}^{-1}$ since the matrices $f_{1}^{-1}$ and $f_{2}^{-1}$ arising from Eq.~(\ref{eq:exfaccon}) are related by the ADHM constraints (the equality $f_{1}^{-1}=f_{2}^{-1}$ implied by Eq.~(\ref{eq:exfaccon}) reproduces two of the original $U(N)$ $k=2$ ADHM constraints). Upon inverting either $f_{1}^{-1}$ or $f_{2}^{-1}$, it remains to determine $V$ and $U^{\prime}$ using the selected form of $f$. The matrix $V$ in Eq.~(\ref{eq:vroot}) is manifestly Hermitian due to the Hermiticity of $f$. From Eq.~(\ref{eq:vvbar}) the matrix $V^{2}$ can be calculated, yielding an $N \times N$ matrix with entries dependent on the elements of $f$ and $\{u_{1},u_{2}\}$. In order to determine $V$, one can take the square root of the matrix $V^{2}$ by first diagonalising $V^{2}$ and then taking the square root of each element in the resulting diagonal matrix. We denote the generic diagonalised matrix $V$ as
\begin{equation}
V = \left( \! \begin{array}{cccc} \lambda_{1} & 0 & \cdots & 0 \\ 0 & \lambda_{2} & \cdots & 0 \\ \vdots & \vdots & \ddots & \vdots \\ 0 & 0 & \cdots & \lambda_{N}
\end{array} \! \right), 
\label{eq:diagv}
\end{equation}
where the $\{\lambda_{v}\}$ are the square roots of the $N$ characteristic values of the matrix $V^{2}$. In this method for performing the matrix square root in Eq.~(\ref{eq:vroot}), the characteristic equation for $V^{2}$ will in general be a polynomial of degree $N$. Thus this method is restricted to $N \geq 5$ unless an explicit similiarity transformation for diagonalising $V^{2}$ can be found. Alternatively, for $N \geq 5$ one could embed the above $U(N)$ $k=2$ solution with $N=4$ and use appropriate coset element factors (which will be quotients of $U(N)$ by the two-instanton stability group) to specify the $U(N)$ two-instanton solution with $N \geq 5$ and then proceed to determine $V$. This is perhaps the only feasible way in which to construct $U(N)$ two-instanton gauge field configurations with $N \geq 5$. 

Given $V$ as in Eq.~(\ref{eq:diagv}), the matrix $U^{\prime}$ can be determined using Eq.~(\ref{eq:uprim}), which has the following explicit form in terms of matrix muliplication for general $k$:
\begin{equation}
U^{\prime}_{[2k]\times[N]} = -\Delta^{\prime}_{[2k]\times\underline{[k]}\times\underline{[2]}}f_{\underline{[k]}\times\underline{[k]}}\bar{u}_{\underline{[2]}\times\underline{[k]}\times\underline{[N]}}\bar{V}^{-1}_{\underline{[N]}\times[N]}.
\label{eq:uprim2}
\end{equation}
For $k=2$, given the form of $V$ in Eq.~(\ref{eq:diagv}), for generic $N$, the matrix $U^{\prime}$ has the following form from Eq.~(\ref{eq:uprim2}):
\begin{equation}
U^{\prime}_{[4]\times[N]} = - \left( \! \begin{array}{cccc} 
U^{\prime}_{11} & U^{\prime}_{12} & \cdots & U^{\prime}_{1N} \\
U^{\prime}_{21} & U^{\prime}_{22} & \cdots & U^{\prime}_{2N} \\
U^{\prime}_{31} & U^{\prime}_{32} & \cdots & U^{\prime}_{3N} \\
U^{\prime}_{41} & U^{\prime}_{42} & \cdots & U^{\prime}_{4N} \\
\end{array} \! \right),
\label{eq:uprim4}
\end{equation}
where we have utilised the definitions in Eq.~(\ref{eq:newvar}) and have written the elements of $f$ as $f_{ij}$. The elements of $U^{\prime}$ in Eq.~(\ref{eq:uprim4}) have the form:
\begin{eqnarray*}
U^{\prime}_{1v} & = & \frac{1}{\lambda_{v}}[A_{1}(\bar{u}_{v1,1}f_{11}+\bar{u}_{v2,1}f_{12})+b(\bar{u}_{v1,1}f_{21}+\bar{u}_{v2,1}f_{22})+\bar{u}_{v1,2}B_{1}+\bar{u}_{v2,2}\beta], \\
U^{\prime}_{2v} & = & \frac{1}{\lambda_{v}}[c(\bar{u}_{v1,1}f_{11}+\bar{u}_{v2,1}f_{12})+A_{2}(\bar{u}_{v1,1}f_{21}+\bar{u}_{v2,1}f_{22})+\bar{u}_{v1,2}\gamma+\bar{u}_{v2,2}B_{2}], \\ 
U^{\prime}_{3v} & = &
\frac{1}{\lambda_{v}}[-\bar{B}_{1}(\bar{u}_{v1,1}f_{11}+\bar{u}_{v2,1}f_{12})-\bar{\gamma}(\bar{u}_{v1,1}f_{21}+\bar{u}_{v2,1}f_{22})+\bar{u}_{v1,2}\bar{A}_{1}+\bar{u}_{v2,2}\bar{c}], \\
U^{\prime}_{4v} & = &
\frac{1}{\lambda_{v}}[-\bar{\beta}(\bar{u}_{v1,1}f_{11}+\bar{u}_{v2,1}f_{12})-\bar{B}_{2}(\bar{u}_{v1,1}f_{21}+\bar{u}_{v2,1}f_{22})+\bar{u}_{v1,2}\bar{b}+\bar{u}_{v2,2}\bar{A}_{2}].
\end{eqnarray*}
The ADHM matrix $U$ for $U(N)$ and $k=2$ is then given by
\begin{equation}
U_{[N+4]\times[N]} = \left( \! \begin{array}{c} V_{[N]\times[N]} \\ U^{\prime}_{[4]\times[N]} \end{array} \! \right), 
\end{equation}
and the corresponding instanton gauge field configuration $A_{m}$ follows from substituting $U$ into Eq.~(\ref{eq:pureg}).

The $U(N)$ two-instanton configuration presented here could conceivably be used in instanton calculus in a number of applications. In particular, testing the proposed exact solutions of $\mathcal{N}=2$ and $\mathcal{N}=4$ supersymmetric gauge theories via the supersymmetric multi-instanton calculus comprehensively developed in \cite{Dorey:1996hu,Dorey:1998xb} is now closer to being achieved at the two-instanton level. Thus the one-instanton test of the proposed exact results for the prepotential in $\mathcal{N}=2$ supersymmetric $SU(N)$ gauge theories in \cite{Khoze:1998gy} could be extended to the two-instanton level by constructing the $SU(N)$ extension of the $SU(2)$ instanton measure given in \cite{Osborn:1981yf}. 
Given the exact form of the two-instanton contribution to the prepotential in $\mathcal{N}=2$ supersymmetric $SU(N)$ Yang-Mills theory (with or without matter hypermultiplets), the matching of predictions from instanton calculus to those of the proposed exact results in these theories can be carried out via the scheme proposed in \cite{Argyres:2000ty} at the two-instanton level. This matching has already been performed at the one-instanton level in \cite{Pomeroy:2001rm}, where agreement between the two sets of results was found to be possible for general values of $N$. Another potential application of the solution exists in the context of instanton calculus in $\mathcal{N}=4$ supersymmetric gauge theories \cite{Dorey:1999pd}. 

The method by which the $U(N)$ $k=2$ ADHM constraints were solved has been applied to the $U(N)$ $k=3$ ADHM constraints, but a solution for the $k=3$ case could not be found using it. The larger number of coupled constraints to be solved serve to make the problem much more involved. It appears that progress in determining instanton configurations with higher topological charge via the ADHM construction is severely limited due to the complexity of the ADHM constraints. The exact, finite action solutions of pure Yang-Mills gauge theory which instanton gauge field configurations represent are given implicitly by the ADHM construction for all simple compact Lie groups and topological charge, but one is apparently prevented from explicitly constructing these solutions beyond a topological charge of three due to the inherently complex nature of the ADHM constraints. From the perspective of thoroughly understanding gauge theories, one hopes that this barrier is not absolute. \\
\\

{\Large{\textbf{Acknowledgement}}} \\

The author thanks V.\ V.\ Khoze for advice and many helpful discussions. The author acknowledges receipt of a PPARC studentship. 
\\
\\

{\Large{\textbf{Appendix}}} \\

Here we give the details of the $U(2)$ transformation which can be used to set $U_{x}=0$ and $u_{1,11}=0$ within the ADHM data for the $U(N)$ $k=2$ ADHM instanton. 
Using the isomorphism $U(N) \simeq U(N-1) \times SU(N)$, we can take $\Omega \in U(2)$ to be the product of a $U(1)$ transformation and an $SU(2)$ transformation. Then the $U(1)$ factor of $\Omega$ acts trivially on the sub-matrix $a^{\prime}$, as is evident from Eq.~(\ref{eq:aux}). 
However, the $U(1)$ factor of $\Omega$ acts non-trivally upon the sub-matrices $u_{\alpha}$. Writing $\Omega$ as $\Omega \simeq \Upsilon \times \Xi$, the following $U(1)$ and  $SU(2)$ elements, $\Upsilon$ and $\Xi$, respectively, can be chosen in order to set $U_{x} = 0$ and $u_{1,11} = 0$:
\begin{eqnarray*}
\Upsilon & = & e^{i\chi}1_{[2]\times[2]} \ \in U(1), \\
\Xi & = & \left( \! \begin{array}{cc} \rho e^{i\theta} & \sqrt{1-\rho^{2}}e^{i\phi} \\
-\sqrt{1-\rho^{2}}e^{-i\phi} & \rho e^{-i\theta} \end{array} \! \right) \in SU(2),
\end{eqnarray*}
with
\begin{eqnarray*}
e^{i\chi} & = & \textup{Im}(u_{1,11}), \\ \nonumber \\
e^{i\theta} & = & \sqrt{Q} + i \sqrt{1 - Q}, \\
Q & \equiv & \left[ \frac{\displaystyle{\textup{Im}(U_{z})}}{\displaystyle{\textup{Im}(U_{y})}}+1 \right] \cdot \left[ \frac{\displaystyle{\rho^{2}|u_{1,11}|^{2}}}{\displaystyle{(\textup{Re}(u_{1,12}))^{2}(1-\rho^2)}} + \frac{\displaystyle{2 \, \textup{Im}(U_{y})}}{\displaystyle{\textup{Im}(U_{z})}} \right]^{-1}, \\ \nonumber \\
e^{-2i\phi} & = & -\frac{\displaystyle{\textup{Im}(U_{z})}}{\displaystyle{\textup{Im}(U_{y})}}e^{2i\theta}, \\ \nonumber \\
\rho^2 & = &  \tfrac{1}{2} \pm \sqrt{1 - \frac{\displaystyle{4 \, \textup{Im}(U_{y})\textup{Im}(U_{z})|U_{x}|}}{\displaystyle{\bigl[4 \, \textup{Im}(U_{y})\textup{Im}(U_{z})|U_{x}|-(\textup{Re}(U_{y})\textup{Im}(U_{z})- \textup{Im}(U_{y})\textup{Re}(U_{z}))^{2} \bigr]}}}.
\end{eqnarray*}
We note that the $SU(2)$ part of the $U(2)$ symmetry would also enable one to set $\bar{U}_{y} = U_{z}$, but this choice would have rendered the solution (Eqs.~(\ref{eq:un2sol1}--\ref{eq:trace2})) of the ADHM constraints singular, since if $\bar{U}_{y} = U_{z}$ then $(|u|^{2}-1)^{-1} \rightarrow \infty$. To obtain a physically meaningful solution we are thus induced to choose $U_{x} = 0$. Other $U(2)$ transformations can be implemented to act upon the $8N+4$-parameter $U(N)$ two-instanton solution given in Eqs.~(\ref{eq:un2sol1}--\ref{eq:trace2}).

\end{document}